\documentclass[11pt,print,amsmath,amssymb,aps]{revtex4-2}
\usepackage{CJK}

\usepackage{graphicx}% Include figure files
\usepackage{dcolumn}% Align table columns on decimal point
\usepackage{bm}% bold math
\usepackage[colorlinks, linkcolor=blue]{hyperref}
\usepackage{subcaption}
\usepackage{amsmath}
\usepackage{amsthm,amssymb,bbold}
\usepackage{array} % for better arrays (eg matrices) in maths
\usepackage[dvipsnames]{xcolor}

\begin{document}
\begin{CJK*}{GB}{} % Use default fonts from CJK (see below)

\title{Vortex shedding patterns in holographic superfluids at finite temperature}
\author{Peng Yang}
\email{pengyang23@sjtu.edu.cn}
\affiliation{Wilczek Quantum Center, School of Physics and Astronomy, Shanghai Jiao Tong University, Shanghai 200240, China}
\affiliation{Shanghai Research Center for Quantum Sciences, Shanghai 201315, China}
\author{Shanquan Lan}
\email{ Corresponding author: lansq@lingnan.edu.cn}
\affiliation{Department of Physics, Lingnan Normal University, Zhanjiang 524048, China}
\author{Yu Tian}
\email{ Corresponding author: ytian@ucas.ac.cn}
\affiliation{School of Physical Sciences, University of Chinese Academy of Sciences, Beijing 100049, China}
\affiliation{Institute of Theoretical Physics, Chinese Academy of Sciences, Beijing 100190, China}
\author{Yu-Kun Yan}
\email{ Corresponding author: yanyukun20@mails.ucas.ac.cn}
\affiliation{School of Physical Sciences, University of Chinese Academy of Sciences, Beijing 100049, China}
\author{Hongbao Zhang}
\email{ Corresponding author: hongbaozhang@bnu.edu.cn}
\affiliation{School of Physics and Astronomy, Beijing Normal University, Beijing 100875, China}
\affiliation{Key Laboratory of Multiscale Spin Physics, Ministry of Education,
Beijing Normal University, Beijing 100875, China}

\begin{abstract}

The dynamics of superfluid systems exhibit significant similarities to their classical counterparts, particularly in the phenomenon of vortex shedding triggered by a moving obstacle. In such systems, the universal behavior of shedding patterns can be classified using the classical concept of the Reynolds number $Re=\frac{v \sigma}{\nu}$ (characteristic length scale $\sigma$, velocity $v$ and viscosity $\nu$), which has been shown to generalize to quantum systems at absolute zero temperature. However, it remains unclear whether this universal behavior holds at finite temperatures, where viscosity arises from two distinct sources: thermal excitations and quantum vortex viscosity. Using a holographic model of finite-temperature superfluids, we investigate the vortex shedding patterns and identify two distinct regimes without quantum counterparts: a periodic vortex dipole pattern and a vortex dipole train pattern. By calculating the shedding frequency, Reynolds number, and Strouhal number, we find that these behaviors are qualitatively similar to empirical observations in both classical and quantum counterparts, which imply the robustness of vortex shedding dynamics at finite-temperature superfluid systems.

\end{abstract}

\maketitle
\end{CJK*}
\tableofcontents

\section{introduction}
The superfluid critical velocity, a fundamental property of superfluids, was first characterized following the discovery of superfluidity in helium-4 during the $\lambda$ transition at $2.17 K$ by Kapitza \cite{1938Natur.141...74K} as well as Allen and Misener \cite{1938Natur.142..643A} . Subsequent studies have further explored this phenomenon in diverse systems, including ultracold atomic gases exhibiting Bose-Einstein condensation \cite{PhysRevLett.84.806,2005Natur.435.1047Z}. Superfluids flow without friction as long as their velocity stays below a critical threshold. However, when this velocity is exceeded, normal fluid components emerge within the superfluid, introducing viscosity and disrupting the frictionless flow. Based on a profound understanding of the results of specific heat experiments and a pioneering insight into elementary excitations, Landau \cite{PhysRev.60.356} proposed that the normal fluid part arises from two kinds of thermal excitations, each contributing a distinct form of specific heat
and both should be found in the low energy region of energy spectrum of superfluids. By accounting for all low-energy excitations during the flow of a superfluid, the critical velocity hence is thereby governed by the minimum value of $\frac{\epsilon(\mathbf{p})}{p}$, where $\epsilon(\mathbf{p})$ is the energy of low-energy excitatoin, such as phonon and roton, with momentum $\mathbf{p}$. However, various experiments, particularly those involving helium-4 flowing through a capillary tube, reveal that the measured critical velocity of superfluids is significantly lower than the prediction made by the Landau criterion. These experiments suggest that the critical velocity is not only influenced by the diameter of the capillary tube but is also determined by another type of gapped excitation in the superfluid, namely the quantized vortex. By investigating superfluidity through the lens of low-energy excitations, a variety of intriguing phenomena have been observed, for example, snake instability in the presence of soliton \cite{PhysRevLett.86.2926,doi:10.1126/science.1062527}, roton instability prior to supersolid formation \cite{PhysRevLett.104.195302}, Giant vortices or vortex lattice in rotating superfluids \cite{2020JHEP...02..104L,PhysRevD.107.026006} and the transition to a Berezinskii-Kosterlitz-Thouless phase in two-dimensional systems \cite{1972JETP...34..610B,JMKosterlitz_1973}. Such behaviors highlight the complex interplay between the superfluid's excitation spectrum and external factors, revealing that the system's dynamical response is highly sensitive to both the local environment and boundary conditions. These effects manifest as modifications in the excited energy spectrum of the superfluid, underscoring the need for a deeper understanding of superfluidity, particularly in non-equilibrium dynamics.

\begin{figure}[htb]
\centering
\includegraphics[width=0.4\linewidth]{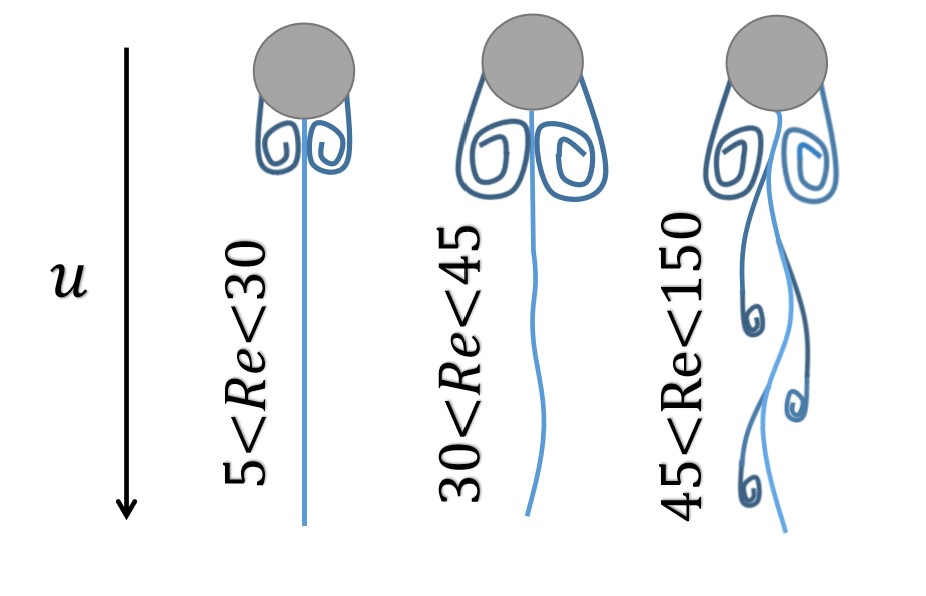}
\caption{\textbf{Classical vortex shedding pattern.} Schematic diagram of classification of classical vortex shedding pattern with different $Re$ in fluid dynamics.}
\label{classical}
\end{figure}

One of the paradigmatic examples explored in the field entails a two-dimensional superflow encountering a cylindrical obstacle, a situation that can precipitate vortex shedding under specific conditions. In this context, the obstacle imposes an internal boundary within the superfluid, inducing significant perturbations in the local superflow velocity. In the rest frame of the superfluid, the obstacle traverses at a constant velocity, engendering a velocity gradient between its leading and trailing edges. As a consequence of the Bernoulli effect, vortices nucleate at regions of maximal local velocity and are subsequently advected into the wake of obstacle, forming a characteristic shedding pattern. The interaction between the generated vortices and the obstacle results in a drag force \cite{PhysRevLett.69.1644,PhysRevLett.82.5186,T_Winiecki_2000,PhysRevA.61.051603,PhysRevA.87.013637,PhysRevA.107.023310,2022EPJP..137.1216W,PhysRevE.102.032217,PINSKER201736}, arising from the transfer of energy associated with vortex nucleation. 
\begin{equation}
    \frac{dE}{dt}=\mathbf{F}_{drag}\cdot \mathbf{v}
\end{equation}
The critical velocity of superfluids is determined by the onset of drag force around the surface of an obstacle \cite{PhysRevA.62.063612,PhysRevA.62.061601,PhysRevA.87.013637,PhysRevA.107.023310,2022EPJP..137.1216W,PhysRevA.93.023640,2024JLTP..215..430K,PhysRevA.109.013317}, which is numerically calculated using the Gross-Pitaevskii equation and corroborated by experimental results \cite{PhysRevA.92.033613,PhysRevLett.121.225301,PhysRevLett.104.160401,PhysRevLett.127.095302}. The dynamics of shed vortices are known to exhibit behavior analogous to the classical B\'enard-von K\'arm\'an vortex street, where a sequence of vortices with alternating signs is periodically shed in the wake. This phenomenon is first observed and investigated in the context of Bose-Einstein condensates by exploring various parameters within the phase diagram of shedding patterns \cite{PhysRevLett.104.150404} and it has been subsequently observed through experiments
 \cite{PhysRevA.92.033613}. In classical fluid dynamics, there is a dimensionless quantity Reynolds number $Re=\frac{v \sigma}{\nu}$ that helps estimate different fluid flow patterns by measuring the quantities of a flow with characteristic length scale $\sigma$, velocity $v$, and kinematic viscosity $\nu$. At low $Re$, a steady and symmetric vortex structure forms behind an obstacle. However, as $Re$ increases beyond a critical threshold, this symmetry is spontaneously broken, leading to the formation of the well-known B\'enard-von K\'arm\'an vortex street \cite{doi:10.1143/JPSJ.11.302} (see Fig.\ref{classical}). However, the situation differs in quantum fluids, where the symmetric shedding pattern is inherently unstable \cite{NORE1993154,PhysRevA.67.023604,PhysRevA.90.013612}, resulting in the exclusive emergence of asymmetric shedding patterns in the phase diagrams of various Bose-Einstein condensate systems \cite{PhysRevA.93.023640,Li_2019,PhysRevE.102.032217,Kai-Hua-Shao60501}.
 
Due to the absence of viscosity in superfluids, the Reynolds number cannot be directly applied to characterize their flow behavior, despite the apparent similarities between classical and quantum fluid dynamics. This limitation was addressed by \cite{PhysRevLett.114.155302}, who numerically investigates the Strouhal-Reynolds relation using the two-dimensional Gross-Pitaevskii equation. Their work, later extended theoretically to superfluid helium-4 by \cite{2015JETPL.102..105S}, has also been employed experimentally to investigate the universality of vortex shedding dynamics \cite{PhysRevLett.117.245301}. \cite{PhysRevLett.114.155302} introduces the concept of effective viscosity in superfluids, which is dynamically generated through the creation of quantized vortices. These vortices possess a quantum of circulation $\kappa=h/m$, where $h$ the Planck's constant and $m$ the mass of atom, giving this quantity the same dimensionality as classical viscosity. Consequently, the superfluid Reynolds number at zero temperature can be defined as
\begin{equation}
    Re_s=\frac{(v-v_c)\sigma}{\kappa}.
\end{equation}

In this paper, we employ a holographic superfluid model \cite{PhysRevLett.101.031601,SeanA.Hartnoll_2008} to explore non-equilibrium vortex shedding dynamics caused by a moving obstacle at finite temperature. The holographic superfluid model, based on the principles of anti-de Sitter/conformal field theory (AdS/CFT) duality \cite{1998AdTMP...2..231M,1999IJTP...38.1113M}, has demonstrated remarkable effectiveness in studying quantum many-body systems, particularly in non-equilibrium systems with strong interactions. Over the past decade, extensive research has focused on understanding non-equilibrium physics through holographic superfluid systems, including Kibble-Zurek mechanism \cite{delCampo:2021rak}, quantum turbulence \cite{doi:10.1126/science.1233529,Du:2014lwa}, dynamical phase transition \cite{PhysRevLett.124.031601} and instability of binary superfluids \cite{An:2024ebg}. Moreover, the universal behaviors displayed during the evolution of superfluids, associated with spontaneously broken symmetries, such as time translation symmetry \cite{PhysRevLett.131.221601}, rotational symmetry \cite{2023JHEP...05..223L,PhysRevLett.131.221602}, and spatial translation symmetry \cite{yang2024cnoidalwavesupersoliditynonequilibrium,gao2024dynamicalbehavioursolitontrain}, have attracted significant interest and may be confirmed through future experiments in cold atom systems. The motivation of our work is to find some universial behaviour of vortex shedding dyanmics at finite temperature. As is well known, the Gross-Pitaevskii model for describing the dynamics of Bose-Einstein condensates is valid only at temperatures close to absolute zero \cite{PhysRevLett.77.420,PhysRevLett.77.988,PhysRevA.56.587}. Therefore, in defining the Reynolds number $Re_s$ for quantum fluids, it is reasonable to replace the viscosity 
$\nu$ with the quantum counterpart $\kappa$. However, the holographic superfluid model operates at finite temperatures and exhibits higher dissipation due to the presence of a normal fluid component \cite{PhysRevD.107.L121901}. In this context, viscosity contributions arise not only from quantum vortices but also from thermal excitations, reflecting the finite-temperature nature of the system. In such a system, does dynamical similarity with classical fluids still exist? More specifically, will the system exhibit a similar vortex shedding pattern, or are there some universal observables that characterizes the finite-temperature superfluid dynamics? In the following section, we will address these questions in detail.

This paper is organized as follwing. In Sec. \ref{2} we introduce the holographic superfluid model used in this study, along with the incorporation of an obstacle modeled using a Gaussian potential. A similar approach with external optical lattices has been employed in \cite{2021JHEP...11..190Y}. Instead of keeping the obstacle moving through the superfluid at a constant speed, we consider a fixed obstacle while the superfluid flows with a constant velocity. In Sec.\ref{3}, we explore vortex shedding pattern in finite-temperature superfluid systems numerically. We examine the critical velocity at which energy dissipation begins and analyze the shedding frequency as a function of the superfluid velocity when it exceeds the critical velocity $v_c$. Additionally, we investigate the behavior of the Strouhal number. In the presence of dissipation, the system tends to form vortex shedding patterns with varying numbers of vortex-antivortex pairs depending on the velocity, a phenomenon less evident in zero-temperature systems. In Sec.\ref{4} we give a summary.

\section{Holographic superfluid model}\label{2}%%%%%%%%%%%%%%%%%%%
The action of holographic superfluid model is given by
\begin{equation}
%\begin{array}{l}Lagrangian
S=\frac{1}{16 \pi G_{4}} \int d^{4} x \sqrt{-g}\left(R+\frac{6}{L^{2}}+\frac{1}{e^{2}} \mathcal{L}_{M}\right),
%\end{array}
\end{equation}
where $G_{4}$ is the gravitational constant in four dimensional spacetime. The first two terms in the parenthesis represent the gravitational part of the Lagrangian, where $R$ the Ricci scalar and $L$ the AdS radius (Hereafter we set $L=1$). The third term corresponds to the matter field contribution to the Lagrangian. By ignoring the backreaction, i.e., taking ${e}\to\infty$ limit, the background spacetime is determined by the vacuum Einstein field equations with a negative cosmological constant. In our consideration, the solution corresponds to a Schwarzschild-AdS black brane with the metric given by
\begin{equation}
%\begin{array}{c}
ds^{2}=\frac{1}{z^{2}}\left(-f(z)dt^{2}+\frac{1}{f(z)}dz^{2}+d\Omega^{2}\right),\qquad f(z)=1-\frac{z^{3}}{z_{H}^{3}}.
%\end{array}
\end{equation}
\textcolor{black}{Here $z_H$ represents the radius of the black brane horizon and} the Hawking temperature of this black brane is given by $T=\frac{3}{4\pi z_{H}}$. In the holographic model, the presence of the black brane leads to an energy flux being absorbed into the horizon during dynamic evolution \cite{doi:10.1126/science.1233529}. \textcolor{black}{As we shall demonstrate, this energy flux serves as a diagnostic tool to identify the onset of vortex shedding}. For numerical simplicity, we choose $z_{H}=1$.

\subsection{Matter field Lagrangian for holographic superfluids}%%%%%%%%%%%%%%%%%%%%%%%%%%%%%%%%%%%%%%%%%%%%%
As proposed in \cite{PhysRevLett.101.031601}, the original hologarphic superfluid matter field Lagrangian density consist of a complex scalar field $\Psi$, a gauge field $A_{\mu}$ and a minimum coupling between them.
\begin{equation}
%\begin{array}{l}
\mathcal{L}_{M}=-\frac{1}{4} F_{\mu \nu}F^{\mu \nu}-\left|D_{\mu} \Psi\right|^{2}-m^{2}|\Psi|^{2},
%\end{array}
\end{equation}
where $D_{\mu}=\partial_{\mu} \Psi-i A_{\mu} \Psi$. The equations of motion for $\Psi$ and $A_{\mu}$ are
\begin{equation}\label{JP}
%\begin{array}{c}
\frac{1}{\sqrt{-g}} D_{\mu}\left(\sqrt{-g} D^{\mu} \Psi\right)-m^{2} \Psi=0,
%\end{array}
\end{equation}
\begin{equation}\label{JA}
%\begin{array}{c}
\frac{1}{\sqrt{-g}} \partial_{\mu}\left(\sqrt{-g} F^{\mu \nu}\right)=i\left(\Psi^{*} D^{\nu} \Psi-\Psi\left(D^{\nu} \Psi\right)^{*}\right).
%\end{array}
\end{equation}
It is straightforward to see that the trivial solution $\Psi=0$ corresponds to normal fluid phase of the syste, where no superfluids condensate exists. As the temperature decreases, a $T_{c}$ is reached. Below $T_{c}$, a nontrivial solution with $\Psi\neq0$ emerges, signaling the spontaneous breaking of $U(1)$ symmetry and the formation of a superfluids condensate. Hereafter, we will focus on the $T<T_{c}$ case, which corresponds to the superfluids system of the boundary field theory.

Based on the asymptotic analysis of bulk fields near the AdS boundary, one obtain
\begin{equation}\label{At}%\nonumber At
A_{t} \sim a_{t}-z^{d-2} \rho+\ldots,
\end{equation}
\begin{equation}%\nonumber Ax
A_{i} \sim a_{i}-z^{d-2} j_{i}+\ldots,  i \in{x,y}
\end{equation}
\begin{equation}%\nonumber psi
\Psi \sim z^{d-\Delta_{+}} \Psi_{0}+z^{d-\Delta_{-}} \Psi_1+\ldots,
\end{equation}
where $\Delta_{\pm}=\frac{d \pm \sqrt{d^{2}+4 m^{2}}}{2}$ with $d=3$ in our case. From holographic dictionary, $a_t$ is chemical potential if there is no external potential,  $\rho$ is the particle number density and $a_i$ is the source of the particle current density $ j_{i}$ along $x$ or $y$ direction. Under the standard quantization, $\Psi_{0}$ is the source of the order parameter of superfluid condensate $\Psi_1$ in the boundary superfluid system. To determine $\Delta_{\pm}$, one convenient choice is $m^{2}=-2$, which will lead to
\begin{equation}%v cos(k_{0}x)$, $v$ is the lattice strength and $k_{0}$ is wave number of the laser
	\Psi \sim z \Psi_{0}+z^{2}\Psi_1+O\left(z^{3}\right)=:z \phi,
\end{equation}
Here we introduce the definition of $\phi$.

\subsection{Critical superfluid velocity without obstacle}
In the absence of obstacle, things can be simplified by choosing the following ansatz
\begin{equation}
\phi=\phi(z), A_t=A_t(z), A_x=A_x(z), A_y=0.
\end{equation}
The corresponding equations of motion of flowing superfluid are 
\begin{equation}%\label{JP}
f\phi'+f'\phi'+\frac{A_t^2}{f}\phi-(A_x^2+z)\phi=0,
\end{equation}
\begin{equation}%\label{JA}
f\phi''-2\phi^2A_t=0,
\end{equation}
\begin{equation}%\label{JA}
fA_x''+f'A_x'-2\phi^2A_x=0.
\end{equation}
The system state with constant particle current can be solved by opening the source $a_x$, in which the superfluid component also has a $U(1)$ current $j_x$ and velocity $v_s$ that defined as
\begin{equation}%\label{JA}
j_x=\frac{i}{2}(\phi D_x\phi^*-c.c.),
\end{equation}
\begin{equation}%\label{JA}
v_s=\frac{j_x}{\frac{i}{2}(\phi D_t\phi^*-c.c.)}.
\end{equation}

\begin{figure}[htb]
\centering
\includegraphics[width=0.45\linewidth]{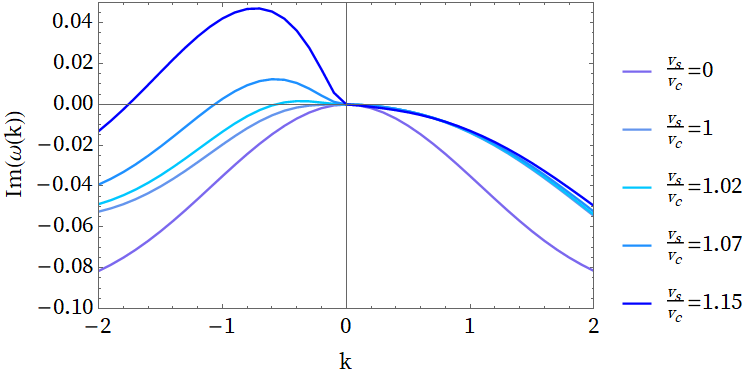}
\includegraphics[width=0.35\linewidth]{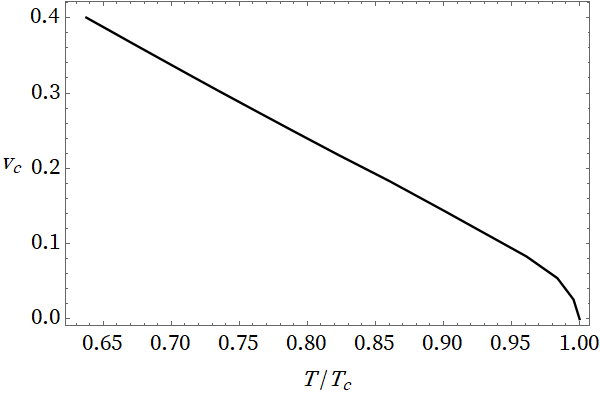}
\caption{\textbf{Dynamical instability and critical velocity of finite temperature supeflow system.} Left: Imaginary part of dispersion relation of sound mode Im$(\omega(k))$ under different superflow background state at temperature $T/T_c=0.712$. The positive Im$(\omega(k))$ of back-propagating sound mode indicates the dynamical instability of superflow state. Right: Critical velocity of superfluid as a function of temperature without obstacle.}
\label{vs}
\end{figure}

When $v_s>v_c$, the superfluid becomes unstable (a phenomenon known as dynamical instability, see Fig.\ref{vs}), which can be analyzed using quasi-normal mode analysis. Assuming the system is in a flowing state
$\{\phi, A_{t}, A_{x}\}$ and subjected to small perturbations, the stable flowing state will exhibit a weak response, with the perturbations exponentially decaying over time. However, in an unstable flowing state, the perturbations will grow exponentially in the linear regime \cite{2021JHEP...11..190Y}. The system's response can be determined by solving the linearized equations of motion for the perturbation fields $\{\delta \tilde \phi e^{-i\omega t+i k x}, \delta \tilde A_{t}e^{-i\omega t+i k x}, \delta \tilde A_{x}e^{-i\omega t+i k x}\}$ and the critical velocity $v_c$ is extracted from emerging positive imaginary part of dispersion relation of sound mode $\omega(k)$. The temperature-dependent suprefluid critical velocity is shown in Fig.\ref{vs}

\subsection{Moving obstacle and comoving transformation}%%%%%%%%%%%%%%%%%%%%%%%%%%%%%%%%%%%%%%%%%%%%%%%
Considering a superfluid system in which an obstacle moves along the $x$ direction with a constant velocity $v_x$ , while the superfluid and normal fluid remain at rest, the potential experienced by the superfluid due to the obstacle can be expressed as $V(x-v_x t,y)$. From the spirit of \cite{2021JHEP...11..190Y}, The moving obstacle can be incorporated into the holographic superfluid model by introducing an external potential term in $a_t$, the temporal component of the gauge field. Specifically, $a_t$ is expressed as
\begin{equation}
    a_t=\mu-V(t,x-v_x t,y),
\end{equation}
where $\mu$ is the constant chemical potential. The response of the superfluid is governed by evolution equations \eqref{JP} and \eqref{JA}, whose explict form in infalling Eddington-Finkelstein coordinates \cite{2021JHEP...11..190Y} are
 %With the choice of the axial gauge $A_z=0$, the above equations of motion can be decomposed into the constraint equation
\begin{eqnarray}
\partial_t \partial_z \phi&=&\partial_z(\frac{f(z)}{2}\partial_z\phi)+\frac{1}{2}\partial^2\phi-i\textbf{A}
\cdot\partial\phi +iA_t\partial_z\phi 
-\frac{i}{2}(\partial\cdot\textbf{A}-\partial_zA_t)\phi
-\frac{1}{2}(z+\textbf{A}^2)\phi, \label{dphi}\\
\partial_t \partial_z \textbf{A}&=&\partial_z(\frac{f(z)}{2}\partial_z\textbf{A})-|\phi|^2\textbf{A}+\textrm{Im}(\phi^*\partial\phi)
+\frac{1}{2}\left[\partial\partial_zA_t+\partial^2\textbf{A} -\partial\partial \cdot\textbf{A}\right], \label{eqeom} \\
\partial_t\partial_z A_t&=&\partial^2A_t-\partial_t\partial\cdot\textbf{A}+f(z)\partial_z\partial\cdot\textbf{A} -2A_t|\phi|^2
+2\textrm{Im}(\phi^*\partial_t\phi)-2f(z)\textrm{Im}(\phi^*\partial_z\phi).
\label{eqAt}
\end{eqnarray}
In addition, there is a constrain equation that should be satisfied
\begin{equation}
0=-\partial_z^2 A_t+\partial_z\partial\cdot\textbf{A}-2\textrm{Im}(\phi^*\partial_z\phi),\label{constrain}
\end{equation}
Here, we choose axial guage $A_z=0$ and $\textbf{A}=(A_x, A_y)$. %Before the obstacle begining to move, we will use Eq.\eqref{dphi}-Eq.\eqref{eqAt} to obtain an initial state, which is a uniform static state with $V=0$. And subsequently, we will turn $V$ on.

 In the next section, we will observe a dynamical transition from a steady state to a vortex-shedding state as the velocity of the obstacle increases. To clearly analyze this transition, we first shift to the co-moving reference frame of the obstacle. A straightforward approach is to consider the superfluid flowing past a stationary obstacle, which appears to be a natural method in the holographic model \cite{yang2024cnoidalwavesupersoliditynonequilibrium}. This can be achieved by adjusting the source term $a_x$ to regulate the particle current density $j_x$. The velocity $v_s$ of superfluid that relatives to obstacle can be obtained gauge invariably $v_s=\frac{\partial_x \theta - a_x}{a_t-\partial_t \theta}$,
where $\theta$ is the phase of $\Psi_1$. However, the presence of dissipation from the thermal bath and quantum viscosity of vortex that created during the evolution of the system will govern the form of 
$\theta(t)$ until the system reaches a steady state, where the velocity $v_s$ becomes smaller than the critical vortex shedding velocity $v_c$. Although it is possible to maintain the velocity by continuously adjusting the source $a_x$, achieving a situation where $v_s<v_c$ remains constant does not seem straightforward. Therefore, in this paper, we present an alternative approach that maintains the obstacle's velocity as relatively constant with respect to the superfluid system. This condition can be interpreted as a balance between external driving forces and internal dissipation, making it more appropriate for our problem.

Since the obstacle's velocity 
$v_x$ remains constant over time, in the co-moving reference frame of the obstacle, the obstacle appears stationary while the superfluid flows continuously with a velocity $v_s=-v_x$. This situation can be modeled through a coordinate transformation $\tilde x=x-v_x t$\footnote{Hereafter, without being confused, we will still use $x$ to express the spatial direction that superfluid flows.}. In the new coordinate system, the obstacle 
$V(x,y)$ becomes time-independent. By keeping the obstacle in motion, the relative velocity $v_x>v_c$ is sustained all the time, regardless of the dissipation strength. This setup effectively mimics the presence of an external force driving the obstacle, which aligns with our objective. The equation of motion in the co-moving coordinate system is provided in the appendix \ref{comove}. In numerical simulation, we set periodic boundary condition in both $x$ and $y$ direction for all fields. The source free boundary condition is given for $\phi$, $A_x$ and $A_y$ in AdS boundary $z_b$ and the total potential is given by $a_t=\mu-V(x,y)$ with obstacle potential taking as Gaussian form
$V(x,y)=A e^{-\frac{x^2+y^2}{2\sigma^2}}$. Hereafter, we fix the size of obstacle with $\sigma=0.025 L_y$ and the height with $A=\mu=8$.

\section{Vortex shedding patterns}\label{3}

\begin{figure}[htb]
\centering
\includegraphics[width=0.3\linewidth]{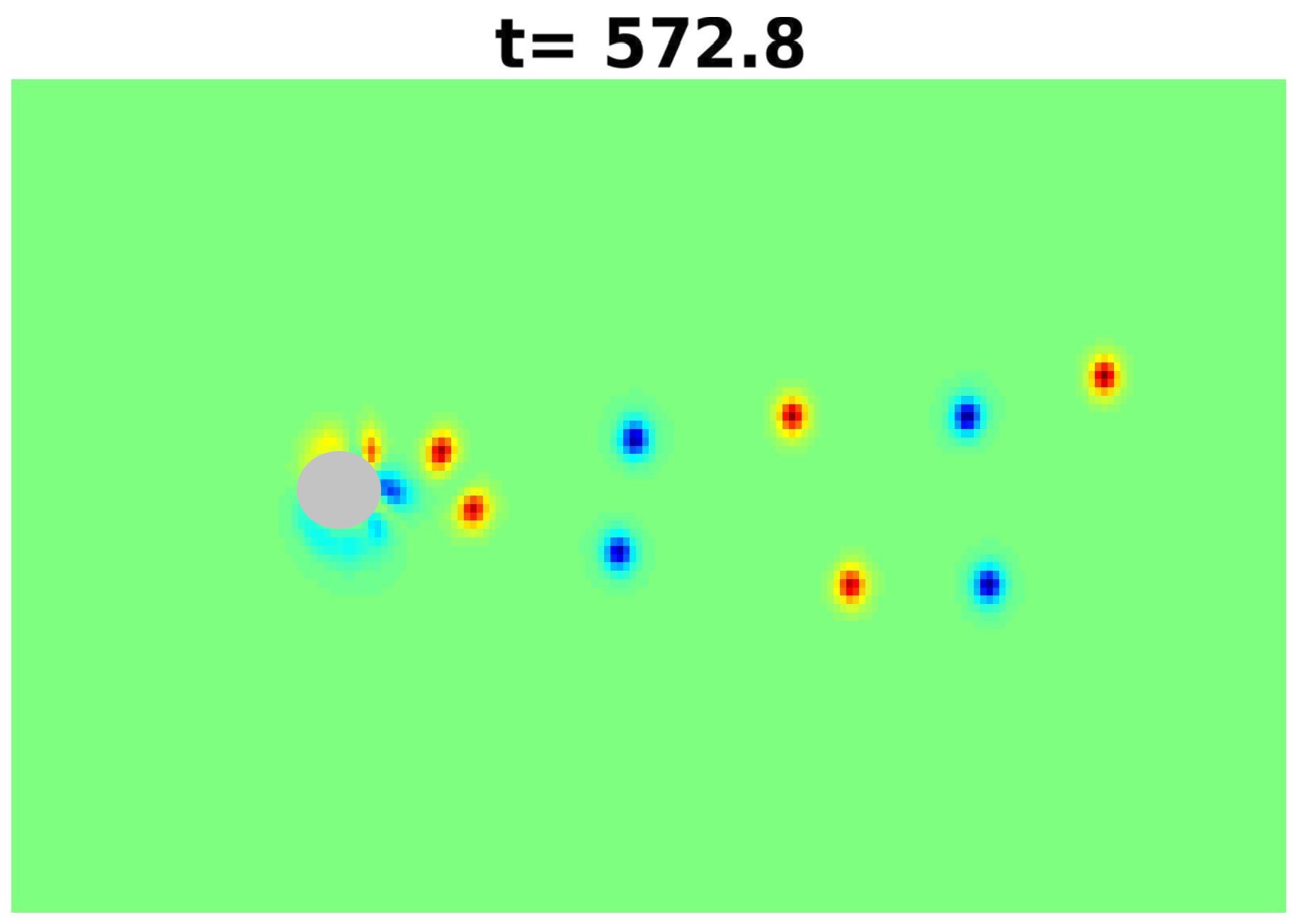}
\includegraphics[width=0.3\linewidth]{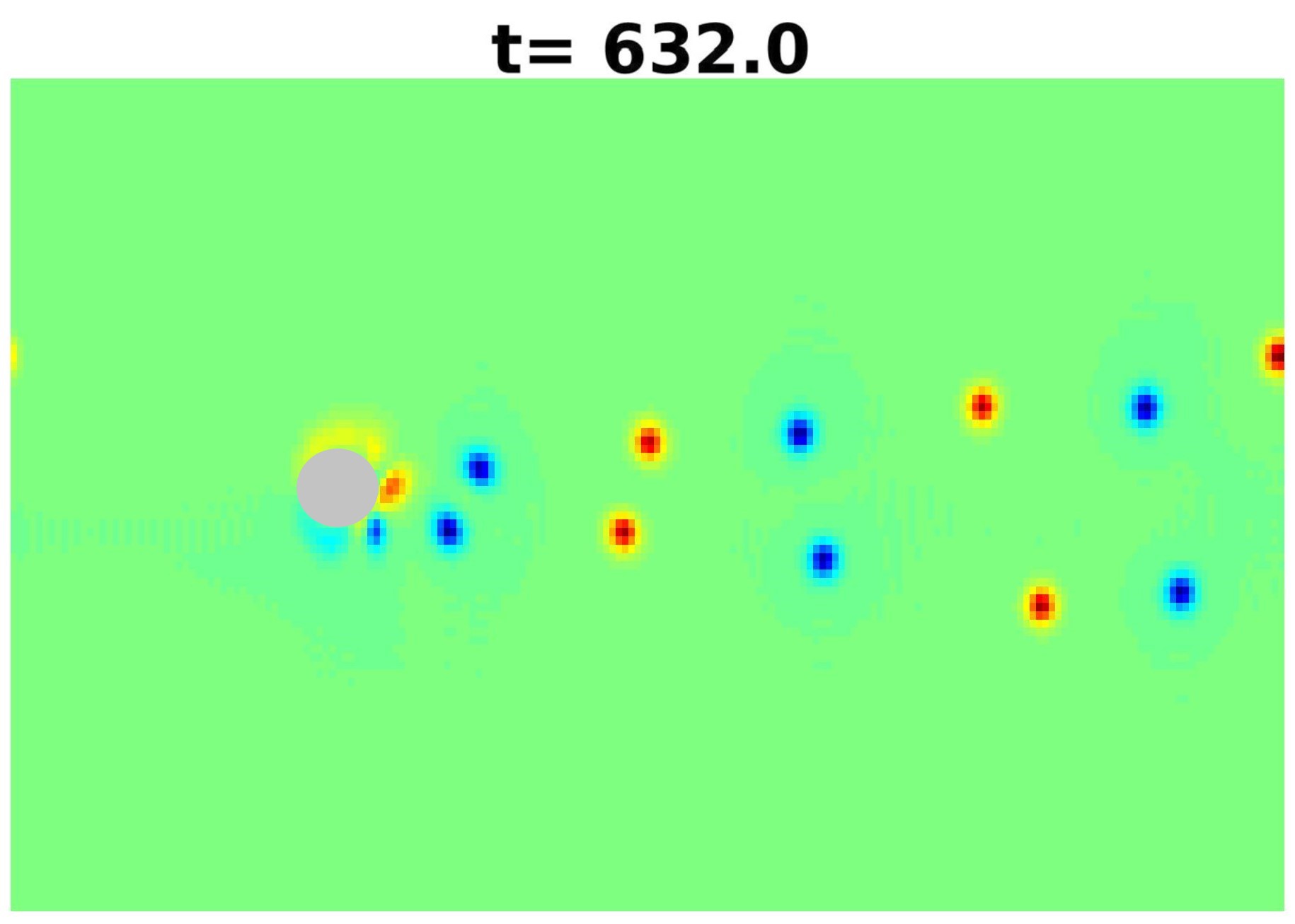}
\includegraphics[width=0.3\linewidth]{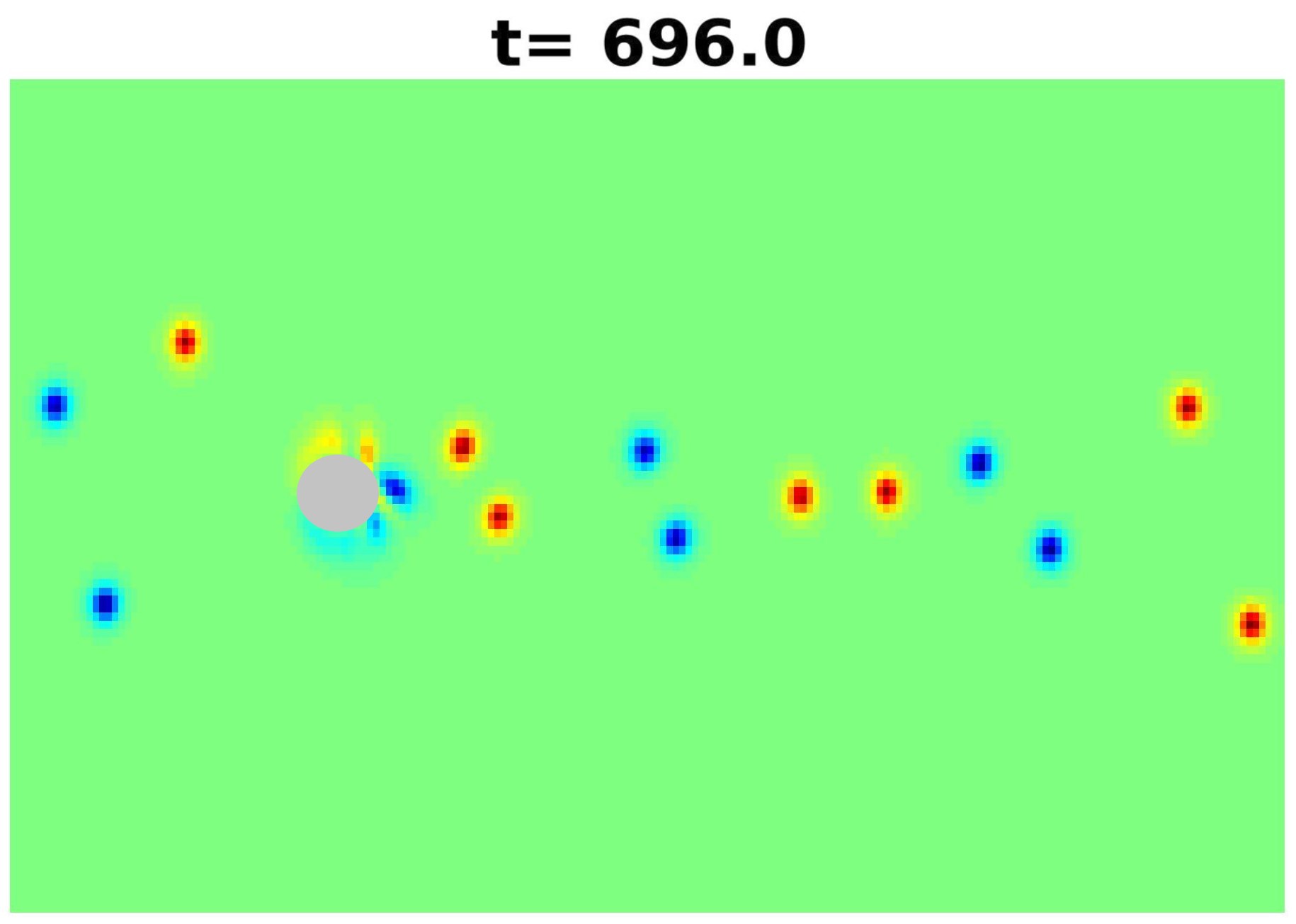}

\includegraphics[width=0.3\linewidth]{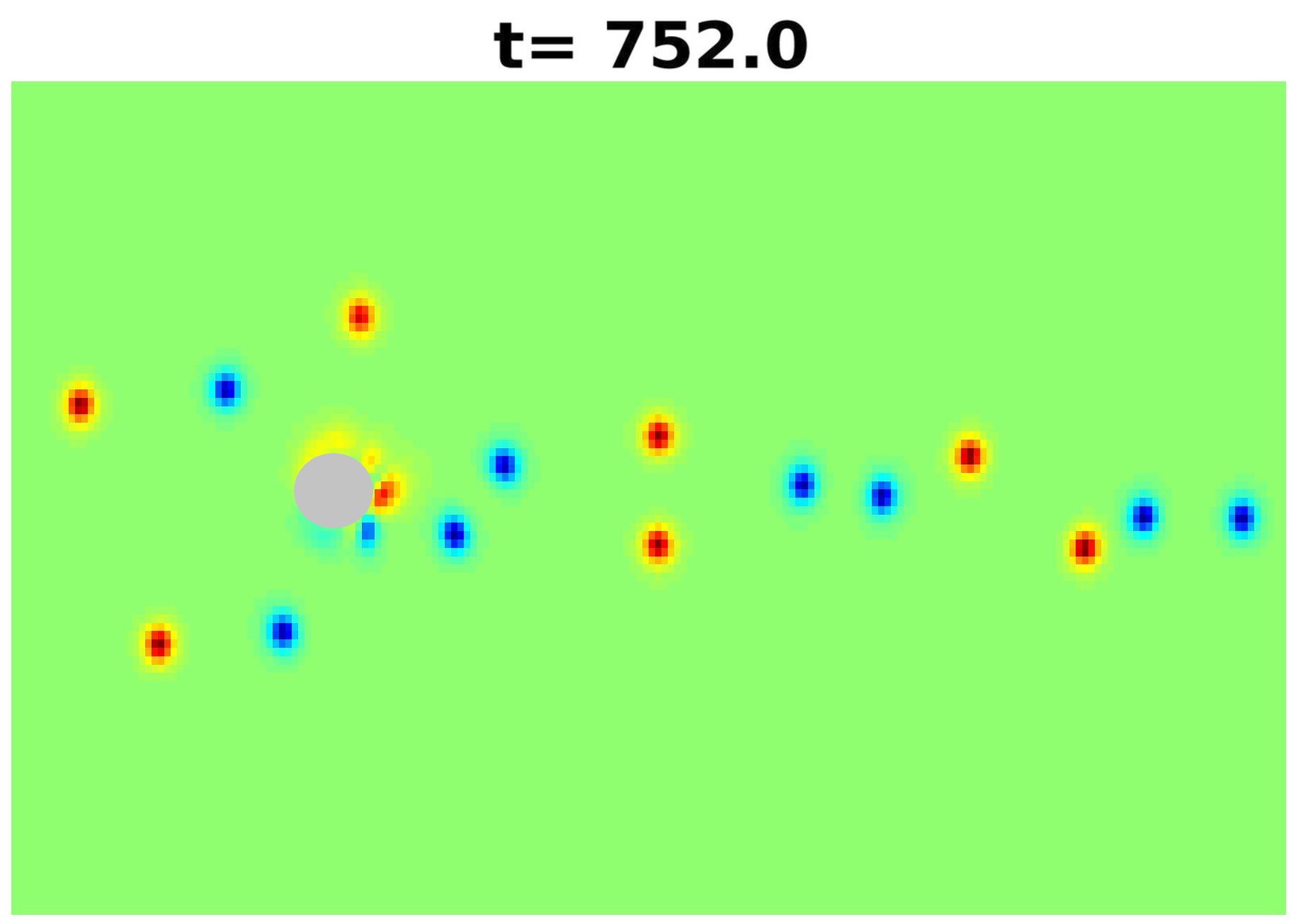}
\includegraphics[width=0.3\linewidth]{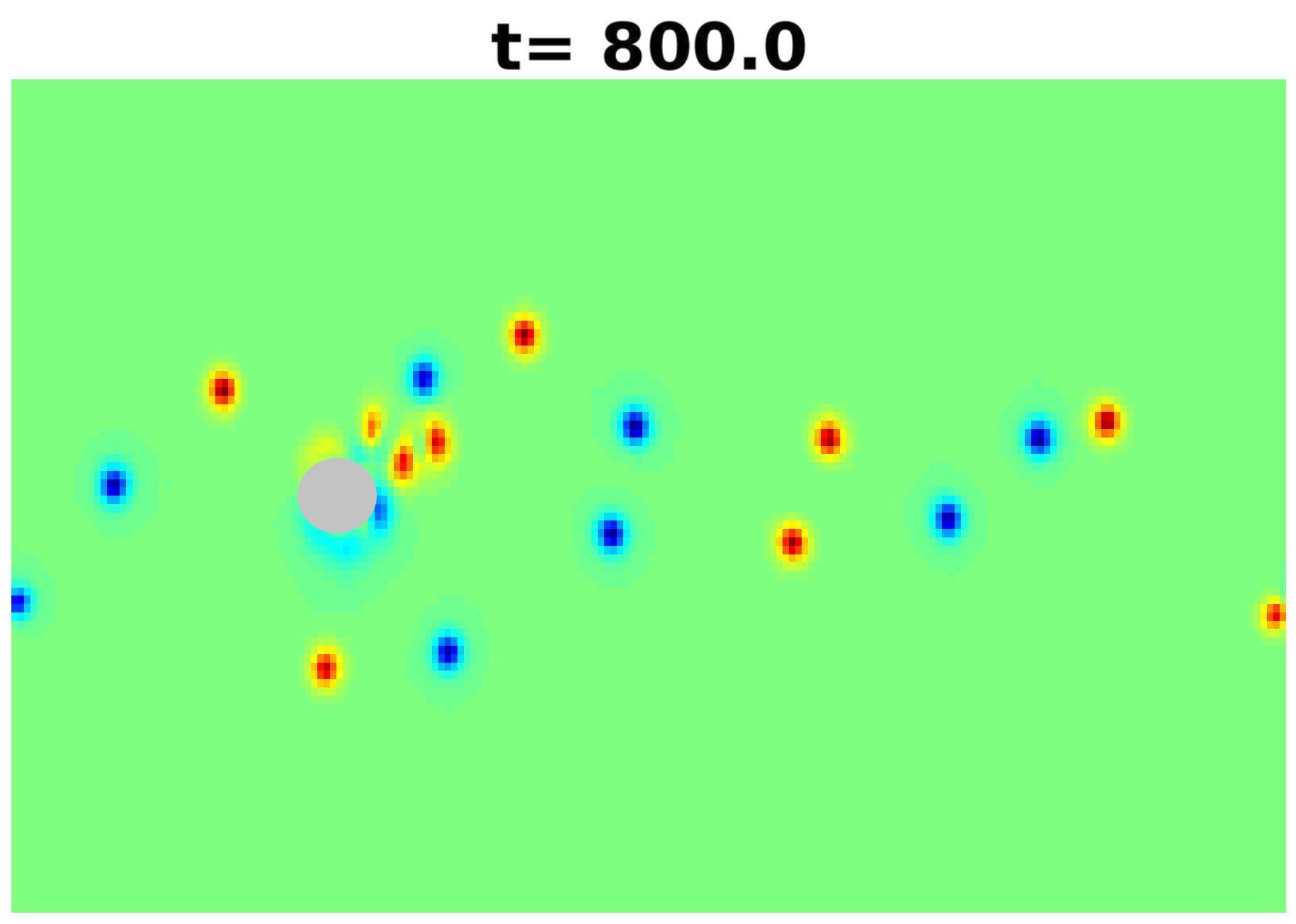}
\includegraphics[width=0.3\linewidth]{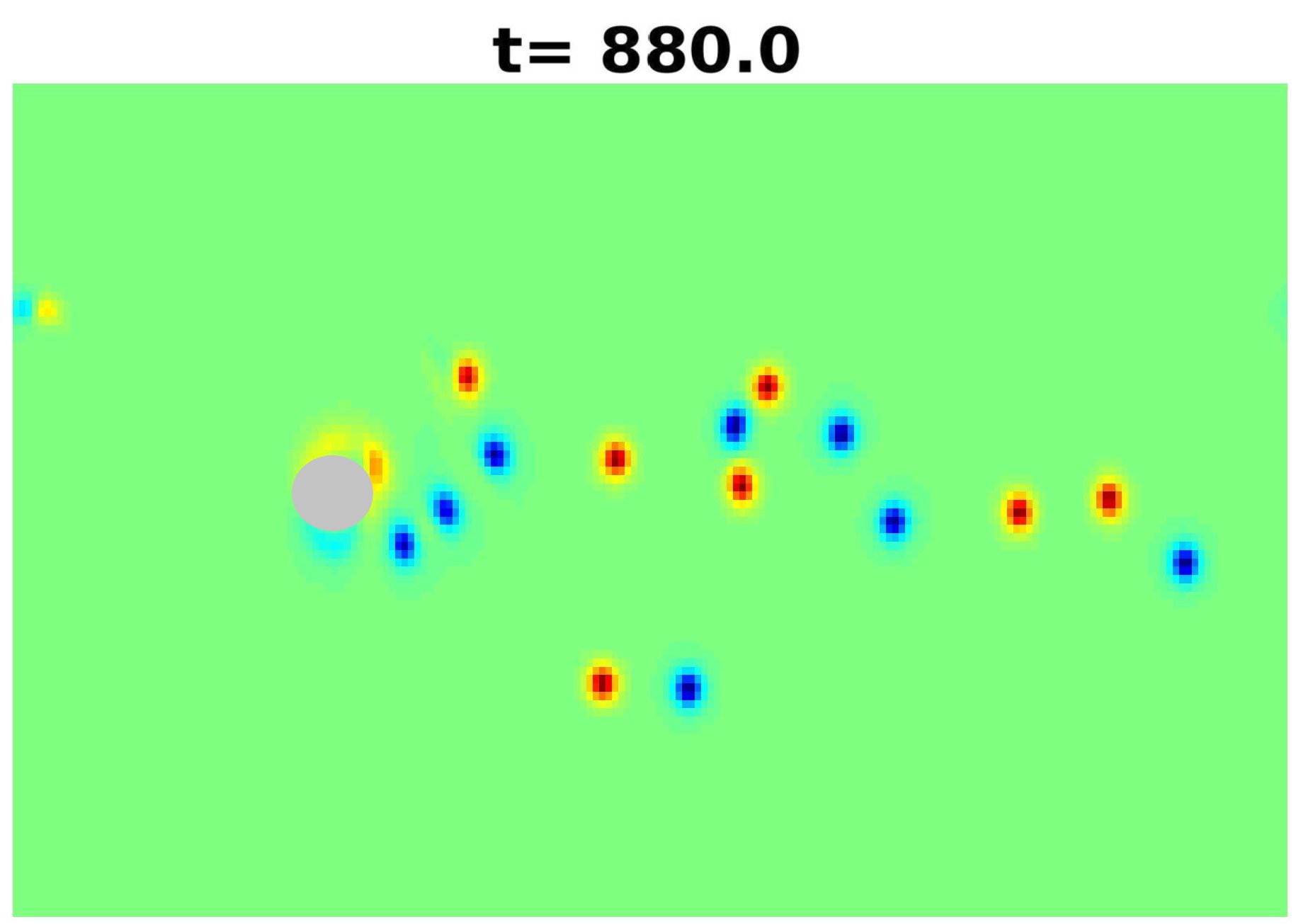}

\includegraphics[width=0.3\linewidth]{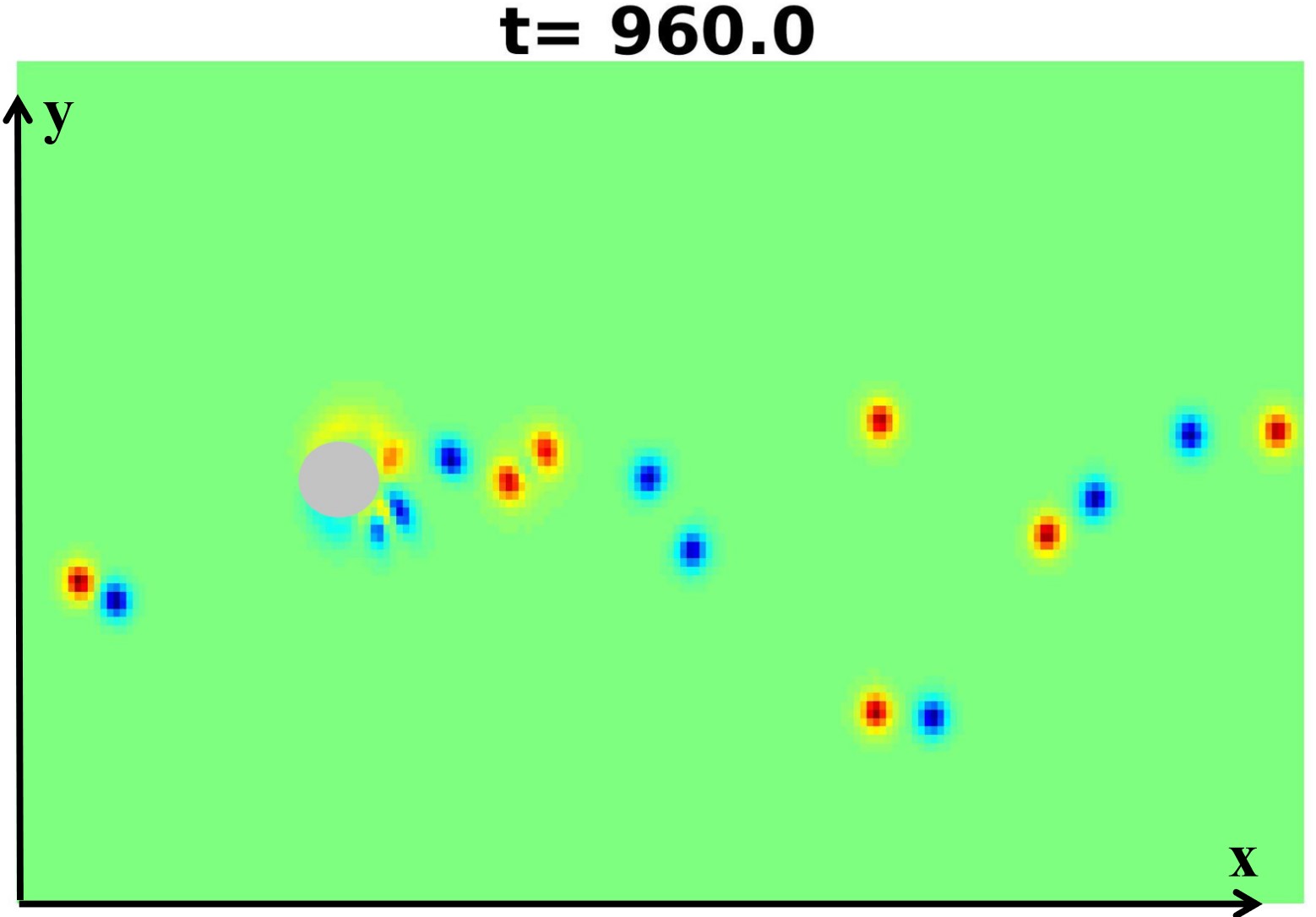}
\includegraphics[width=0.3\linewidth]{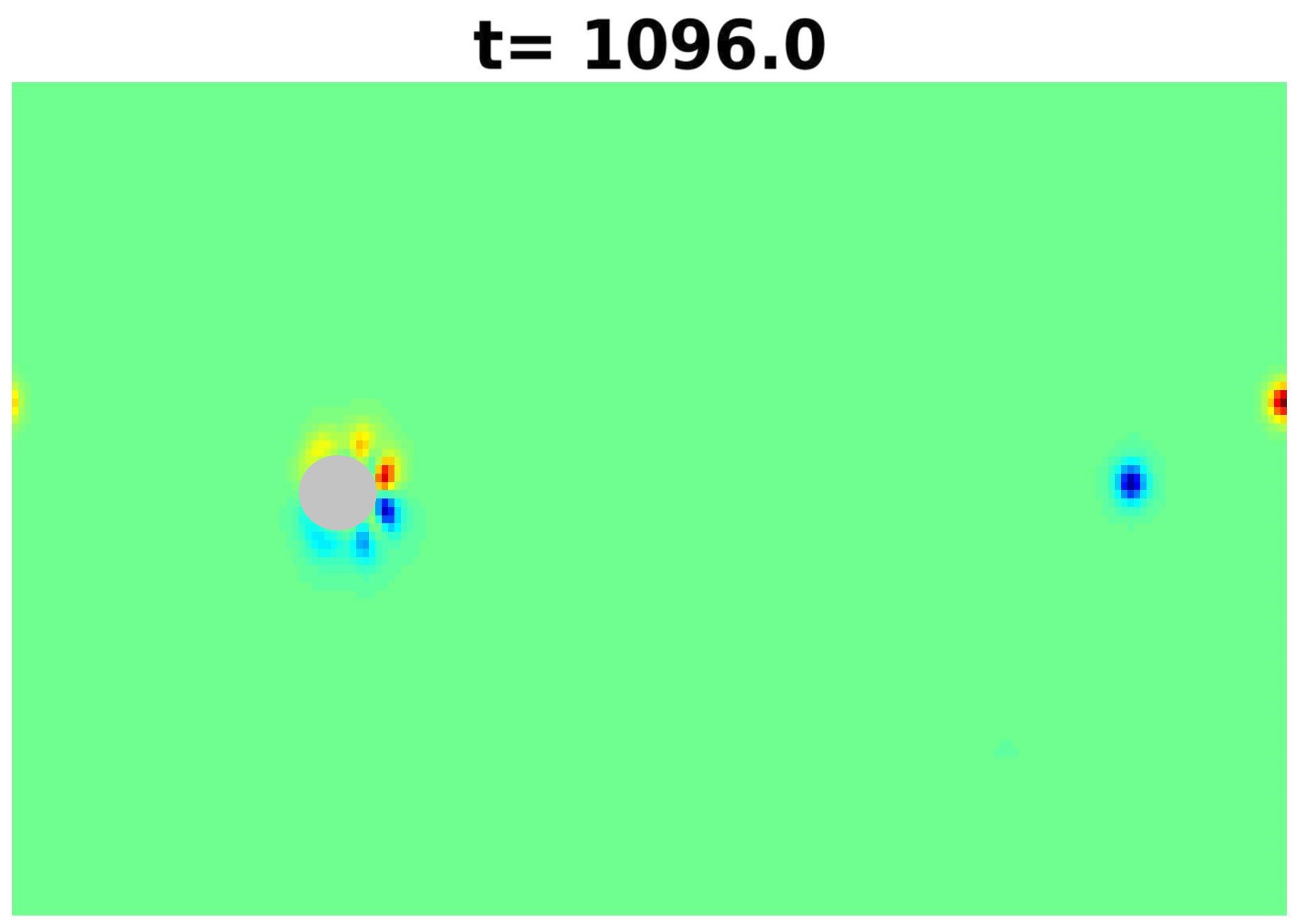}
\includegraphics[width=0.3\linewidth]{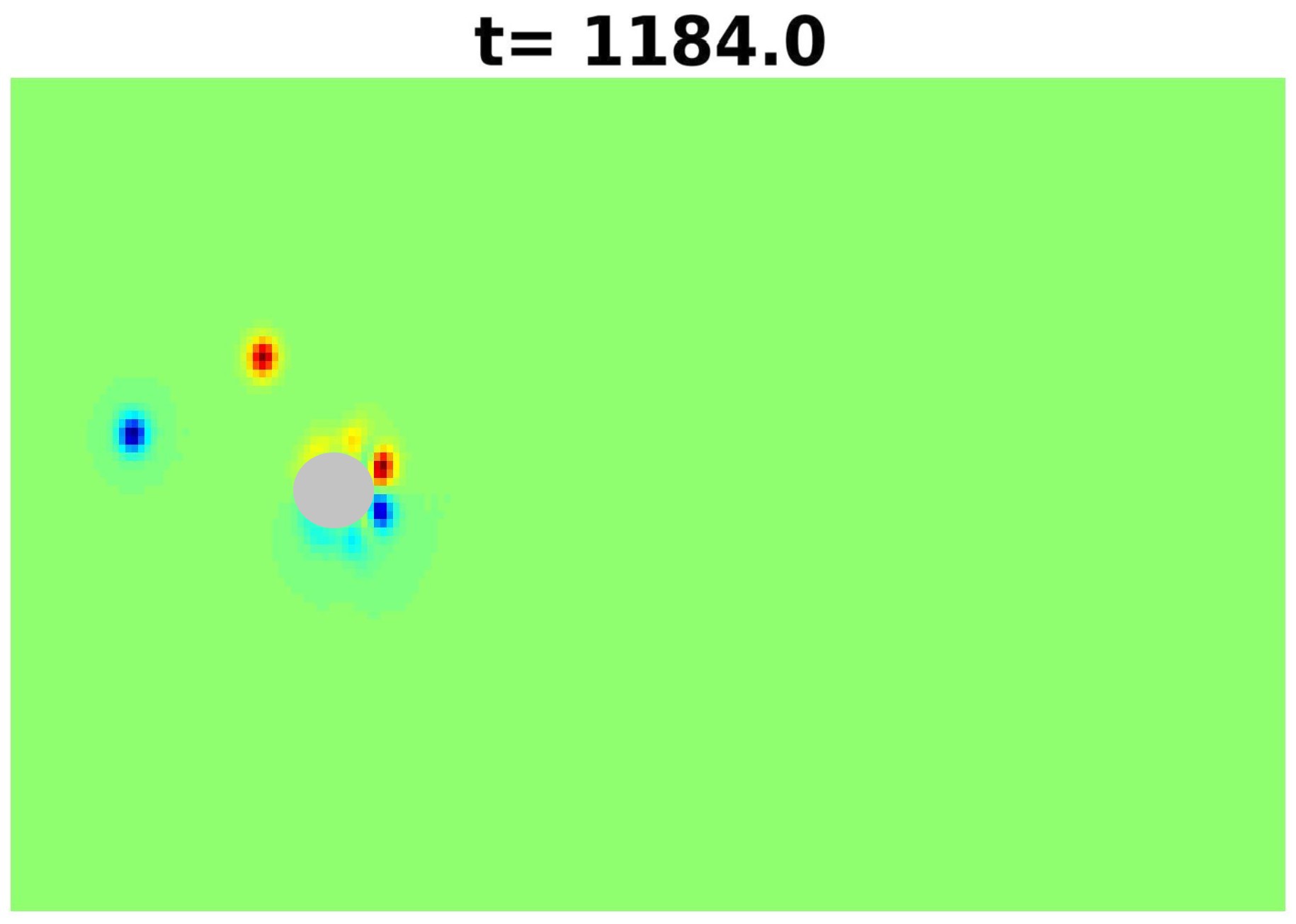}

\caption{\textbf{Time evolution from K$\acute{a}$rm$\acute{a}$n vortex street to periodic vortex shedding.} Since the mirror symmetry is broken, at middle stage, there are a trail of alternating pairs of like-signed vortices being periodically produced after obstacle, which is reminicent of K$\acute{a}$rm$\acute{a}$n vortex street. The effect of dissipation will restrore the mirror symmetry and lead to the late behavior back to periodic vortex shedding pattern. The horizontal axis is parallel to direction of supercurrent and the gray region is the location of obstacle.}
\label{0}
\end{figure}

\subsection{Short-lived K$\acute{a}$rm$\acute{a}$n vortex street to periodic vortex shedding}

To investigate the vortex shedding pattern, we choose the velocity of obstacle is larger than the critical velocity. At the initial time of evolution, we break the spatial mirror symmetry alone the longitudinal direction by introducing small perturbation $\sum \delta f e^{-ik_xx-ik_yy+i\theta}$ with some random phase $\theta$.

\begin{figure}[htb]
\centering
\includegraphics[width=0.3\linewidth]{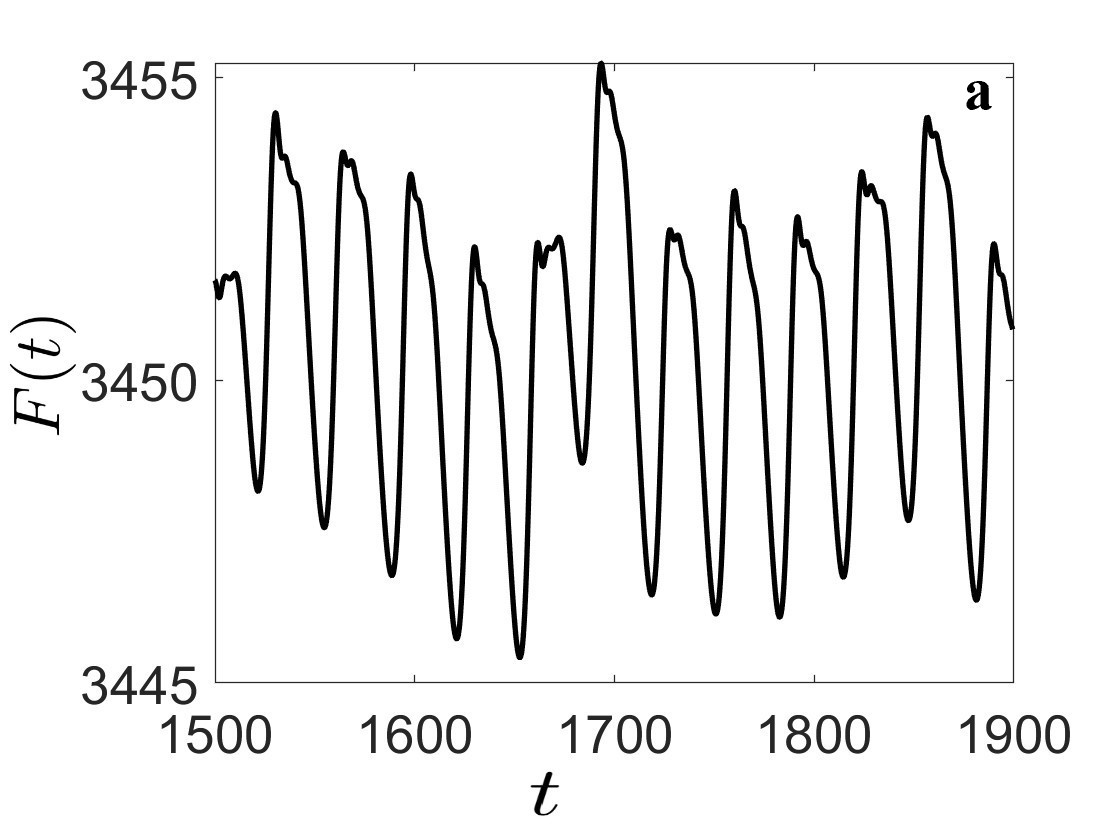}
\includegraphics[width = 0.3\textwidth]{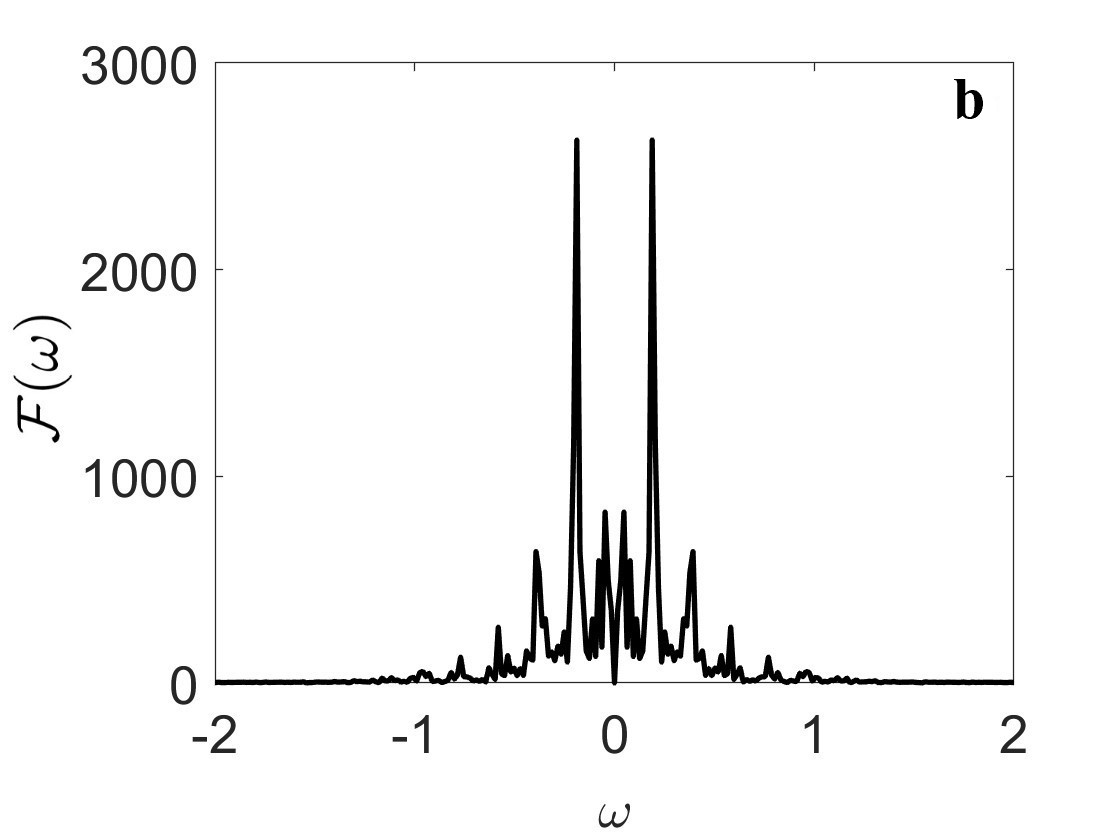}

\includegraphics[width =0.3 \textwidth]{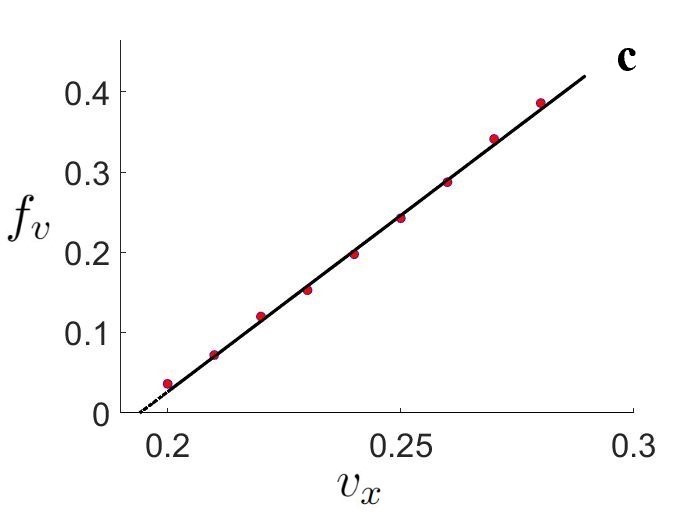}
\includegraphics[width = 0.3\textwidth]{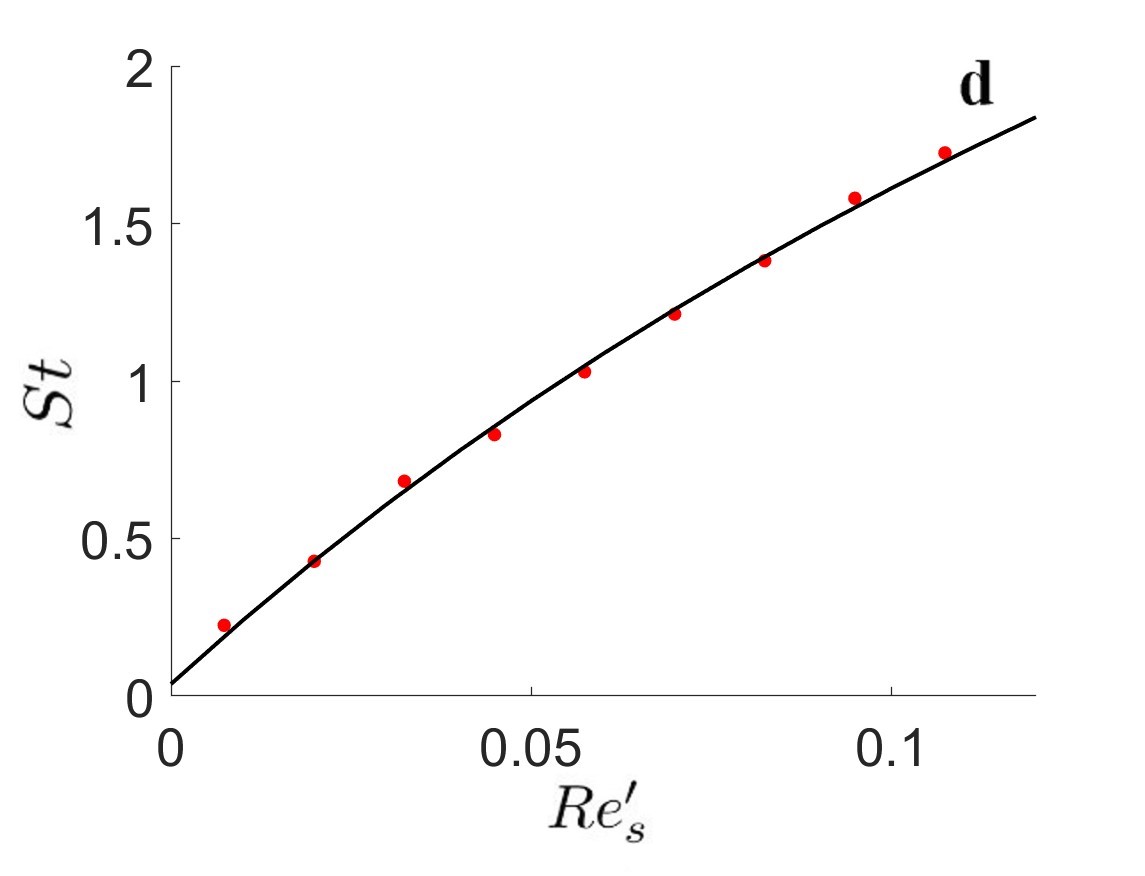}

\caption{\textbf{Late-time behaviour of periodic vortex shedding pattern.} Fig.\ref{FT}a shows time-dependent energy dissipation $F(t)$, which exhibits periodic oscillating behavior over time, corresponding to the periodic shedding of vortex dipoles (The long-wave fluctuation results from the energy fluctuation of vortex-pair annihilation.). Fig.\ref{FT}b shows the Fourier transformation of $F(t)$.Fig.\ref{FT}c shows the frequency of vortex dipole shedding against velocity of obstacle, in which the black fitting line is $f_v=4.4 (v_x-0.1941)$. Fig.\ref{FT}d exhibits Strouhal number as a function of the Reynolds number, the black fitting line  is identical to quantum counterpart $St=st_{\infty}(1-\frac{\alpha}{Re_s'+\beta})$ with $st_{\infty}=6.756$, $\alpha=0.3013,\beta=0.2995$.}
\label{FT}
\end {figure}

In Fig.\ref{0}, we show that there exist a short-lived K$\acute{a}$rm$\acute{a}$n vortex street stage after the perturbation with $v_x=0.24$. However, unlike zero-temperature BEC system where K$\acute{a}$rm$\acute{a}$n vortex street remains stable but the shedding pattern with mirror symmetry is dynamical unstable, in our finite temperature superfluid system, on the contrary, the K$\acute{a}$rm$\acute{a}$n vortex street  persists only for a limited time before the mirror symmetry is restored.

\begin{figure}[htb]
\centering
\includegraphics[width=0.7\linewidth]{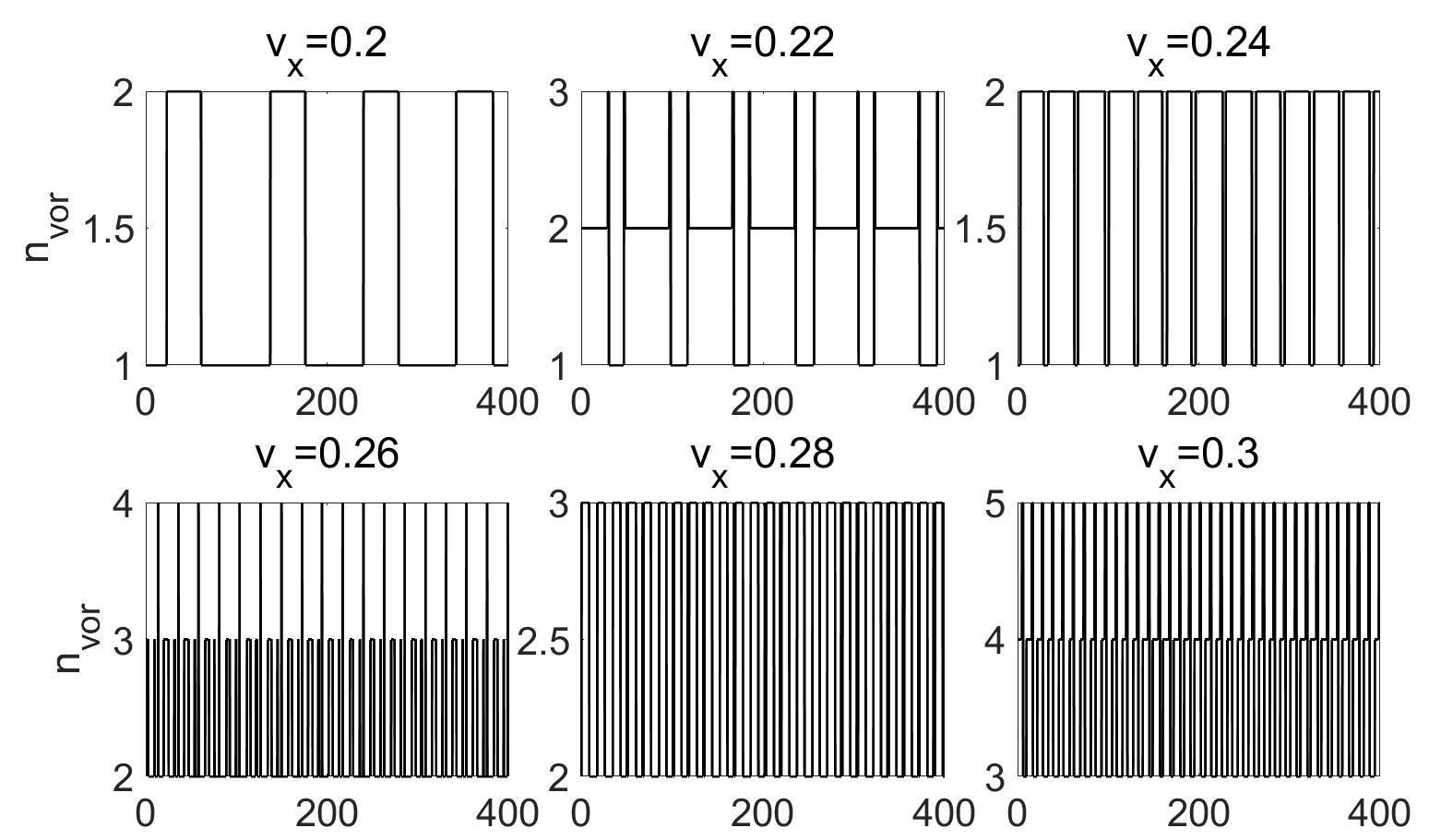}
\includegraphics[width=0.7\linewidth]{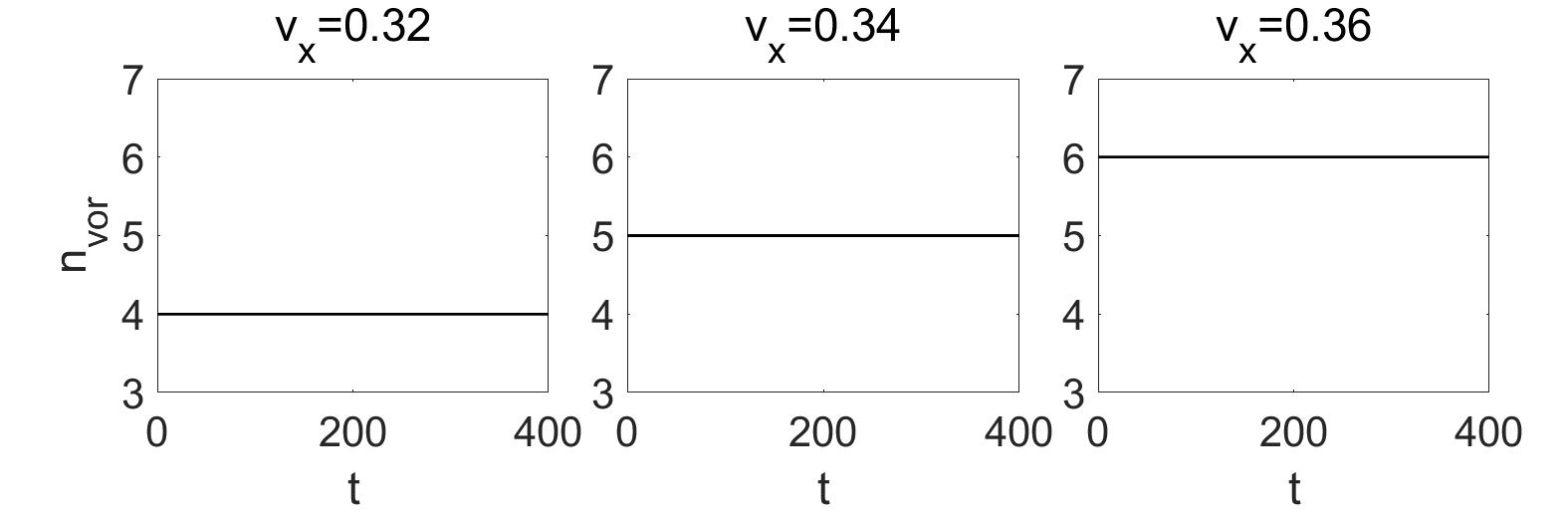}
\caption{\textbf{Number of vortex dipole.} Temporal behavior of number of vortex dipoles in periodic vortex dipole shedding pattern for moving obstacle with $v_x=\{0.2,0.22,$ $0.24,0.26,$ $0.28,0.3,$ $0.32,$ $0.34,$ $0.36\}$, respectively.} 
\label{n_t_0}
\end{figure}

To quantitatively characterize the periodic vortex shedding pattern in the late-time stage that shown in Fig.\ref{0}, we firstly calculate the frequency of vortex dipole shedding. The frequency $f_v$ is determined through the Fourier transformation of time-dependent energy dissipatoin $\mathcal{F}(\omega)$  (see Fig. \ref{dissipation} in App.\ref{comove}). A typical result is shown in Fig.\ref{FT}. We choose the periodic vortex shedding pattern under moving obstacle with $v_x=0.24$ on the background superfluid system at temperature $T /T_c = 0.525$. In this case, we obtain the frequency $f_v=0.189$ from $\mathcal{F}(\omega)$ that shown in Fig.\ref{FT}b. We further calculate the velocity-dependent shedding frequency and display in Fig.\ref{FT}c, from which we can see the the shedding frequency exhibit linear relation with velocity just like the zero-temperature case \cite{PhysRevA.92.033613,Lim_2022}, $f_v=a(v-v_c)$. In this case, the critical velosity $v_c\approx0.194$ that is significantly less than the one without obstacle in Fig.\ref{vs}. Finally, we investigate the similarity of our finite-temperature vortex shedding pattern with quantum fluid flow patterns. By definition, there are two main dimensionless number to characterize the patterns, i.e., Reynolds number $Re_s=\frac{(v_x-v_c)\sigma}{\kappa}$ and Strouhal number $St=\frac{f_v }{v_x}\sigma$. Since in our system, the viscosity is not only coming from vortexs but also comes from normal fluid. One can generalize $\kappa$ to $\eta=\eta(\kappa(T=0),\nu(T>0))$, where $\nu$ is finite-temperature viscosity. The assumaption of constant $\eta$ enable us to redefine $Re_s'=\eta Re_s$. In Fig.\ref{FT}d the relation of $St-Re_s'$ is shown, which also exhibits similarity with zero-temperature behaviour \cite{PhysRevLett.114.155302}.

\subsection{Periodical vortex dipole shedding to steady vortex dipole train}

Unlike in a zero-temperature superfluid system, where created vortex dipoles can travel considerable distances without annihilating, the dissipation emerging at finite temperatures accelerates vortex dipole annihilation. In this section, we examine the number of vortex dipoles shed behind the obstacle. Our findings show that when the obstacle's speed is sufficiently high, the shedding pattern transitions from periodic vortex dipoles shedding to a steady vortex dipole train.

\begin{figure}[htb]
\centering
\includegraphics[width=0.7\linewidth]{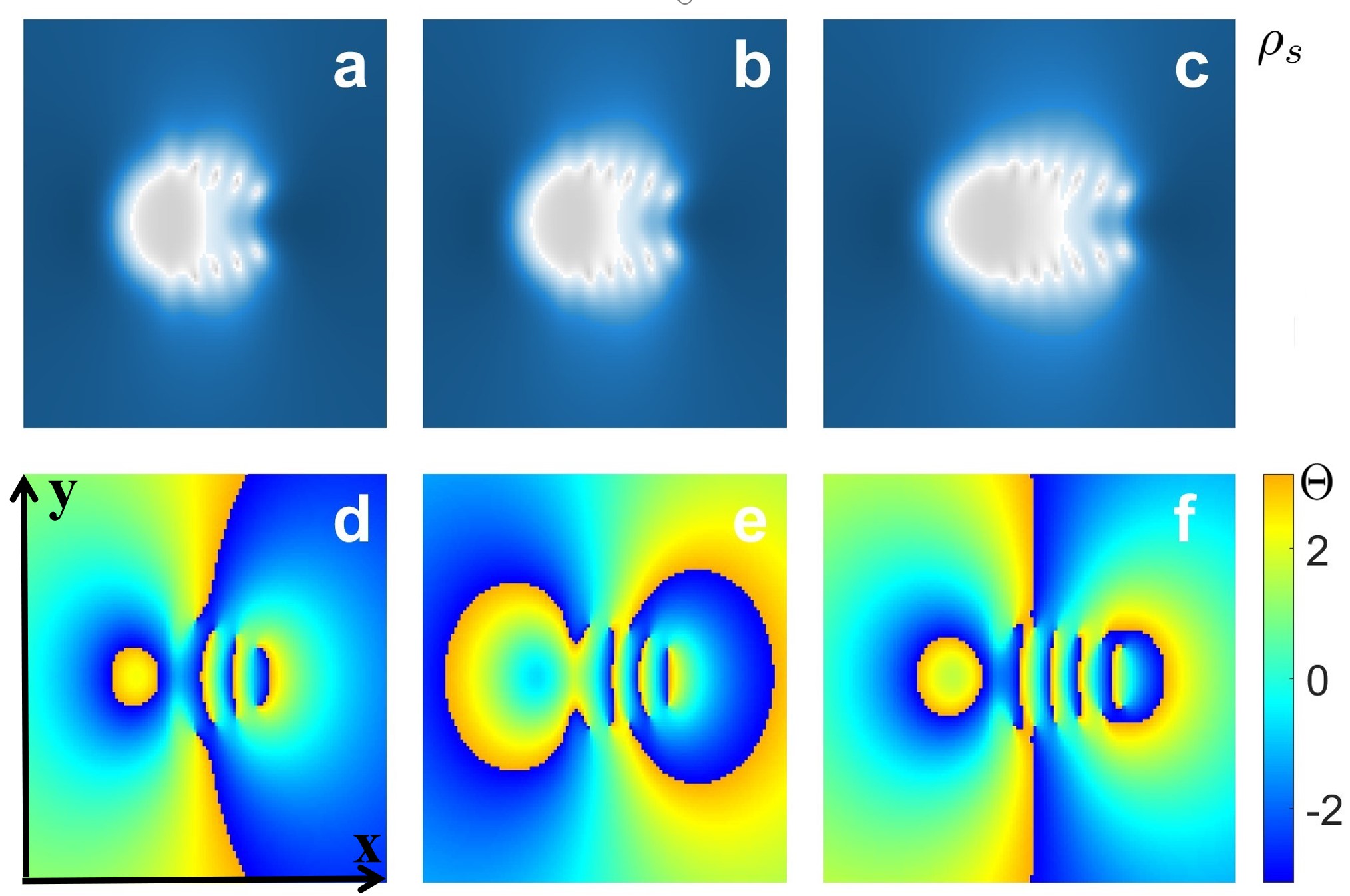}
\caption{\textbf{Superfluids condensate density and phase of order parameter.} Fig.\ref{train}a-Fig.\ref{train}c display the superfluids condensate density $\rho_s(x,y)$ with velocity of obstacle $v_x=\{0.32,0.34,0.36\}$, respectively. The gray region indicates the location of the obstacle, with a train of vortex dipoles trailing in the wake flow. Fig.\ref{train}d-Fig.\ref{train}f exhibit the phase of order parameter $\Theta(x,y)$ with velocity of obstacle $v_x=\{0.32,0.34,0.36\}$, respectively. The location of vortexs are accompanied by phase winding.}
\label{train}
\end{figure}

At low obstacle velocity, we know that there is a periodic vortex dipole shedding pattern. However, since the lifetime of vortex dipole is short, the number of vortex dipole is limited (see Fig.\ref{n_t_0}). For $v_x=0.2$, the vortex dipole is periodically produced and the number of vortex dipoles is also oscillating periodically. Same phenomena is hold for $v_x\in[0.2,0.3]$, but there are two main difference. The first difference is the frequency of vortex shedding, which we have already discuss in detail in the last section (see Fig.\ref{FT}), i.e., the frequency grows linearly with obstacle velocity. And this imformation is well recorded in both  Reynolds number and  Strouhal number. Another difference arises from the mean value of vortex dipole number $ n_{vor}$ (see Fig.\ref{n_t_0}), which indicates how many vortex dipoles can appear simultaneously. A trend is observed, when $v_x$ is increasing the $n_{vor}$ is also increasing. As $v_x$ increasing, one expects that the frequency of vortex dipole shedding becomes large enough such that earlier vortex dipoles do not have time to annihilate before new vortex dipoles are created. Eventually, the shedding pattern transitions from periodic vortex dipoles to a steady vortex dipole train. As shown in Fig.\ref{n_t_0} with velocity lager than $0.3$, the evolution of vortex dipole number is no longer oscillating but becomes constant over time. To see the pattern clearly, we display the superfluids condensate density $\rho_s$ and phase of order parameter $\Theta$ in Fig.\ref{train}, from which we can see that the vortex dipole train appear in wake flow of obstacle.

\begin{figure}[htb]
\centering
\includegraphics[width=0.6\linewidth]{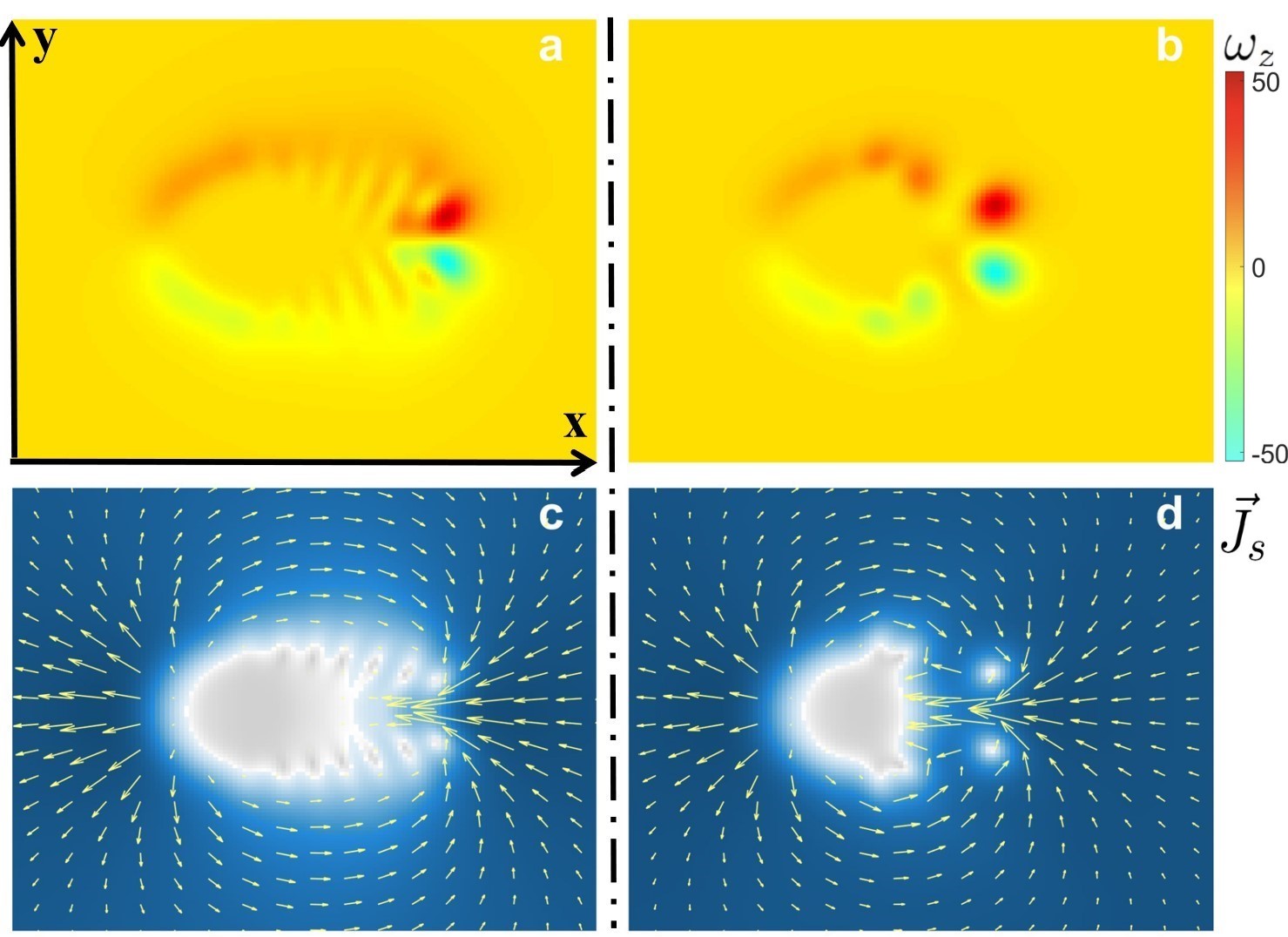}
\caption{\textbf{Vortcity distribution and velocity field vector diagram.} Left column corresponds to the case for obstacle velocity $v_x=0.36$ , while the right column represents the case with obstacle velocity $v_x=0.26$. The upper row shows the Vortcity distribution, while the lower row displays the velocity field vector in the context of the background superfluid condensate density.}
\label{vorticity}
\end{figure}

Given that the static vortex encodes a quantized winding number, the vortex wave function phase takes the form $\Theta=n\theta$ with $\theta$ denoting the angular coordinate and $n=1$ for single positive vortex. The velocity field around the static vortex exhibits a divergent behavior as one approaches the vortex core
\begin{equation}
    \vec{v}_{vor}\sim \frac{\hat\theta}{r}.
\end{equation}
For an adjacent vortex dipole, the velocity fields of themselves interact with each other, leading to a Magnus force that drives the vortices toward annihilation. However, in our system, the vortex dipole train remains static in a co-moving frame of reference, with no annihilation occurring.  To further elucidate the configuration of the vortex dipole train, we investigate the distribution of the supercurrent field $\vec{J}_s$ and the vorticity $\omega_z$ within the static vortex dipole train configuration, the definitions of which are given by the following expressions
\begin{align}
    \mathbf{J_s}=\frac{i}{2}(\Psi_1^*\mathbf{\nabla}\Psi_1-c.c.),\\
    \omega_z=(\mathbf{\nabla\times\mathbf{J_s}})\cdot \hat e_z.
\end{align}

From Fig.\ref{vorticity}a and Fig.\ref{vorticity}b, it is evident that the presence of a non-zero supercurrent results in a vorticity of positive sign for clockwise supercurrents and negative sign for anticlockwise supercurrents. In Fig. \ref{vorticity}b ($v_x=0.26$), the supercurrent attains a local maximum near the vortex positions. In contrast, in Fig.\ref{vorticity}a ($v_x=0.36$), only the earliest vortices significantly influence the vorticity, while the contribution of subsequent vortices diminishes, leading to a reduced overall vorticity. The underlying phenomenon is elucidated by examining the supercurrent field distributions in (Fig.\ref{vorticity}c and Fig.\ref{vorticity}d). For $v_x=0.26$ (Fig.\ref{vorticity}c), the supercurrent field pervades the boundaries of both vortices and the obstacle.  In contrast, for $v_x=0.36$ (Fig.\ref{vorticity}d), the supercurrent is only spread to the peripheries of the vortices and obstacle, with the central regions of the supercurrent field effectively shielded. The underlying reason lies in the fact that the transverse supercurrent is balanced by the array of vortices of like sign, while the longitudinal supercurrent is counterbalanced by the background current. These interplays result in the vortex dipole train behaving as a quasi-rigid entity, just like the obstacle impeding supercurrent penetration, thus maintaining the dipole configuration as a static structure throughout its evolution.

\section{Summary}\label{4}
In this paper, we investigate the behavior of vortex shedding patterns at finite temperatures using a holographic superfluid model. We identify two distinct patterns that do not have quantum counterparts in zero-temperature quantum fluid systems. Unlike the K$\acute{a}$rm$\acute{a}$n  vortex street, which breaks mirror symmetry along the velocity direction, strong dissipation significantly alters the shedding patterns. The K$\acute{a}$rm$\acute{a}$n  vortex street appears only as a short-lived transitional state and ultimately evolves into a symmetric and periodic vortex dipole shedding pattern. Both the shedding frequency and the number of vortex dipoles increase monotonically with the relative velocity between the obstacle and the superfluid. Beyond another critical velocity, the shedding pattern transitions to a steady state, forming a vortex dipole train. Although these dynamical patterns differ from those in classical and quantum systems, we find that empirical relations, such as the shedding frequency with velocity ($f_v-v$) and the Strouhal-Reynolds ($St-Re$) relation, still hold in our simulations and hence  strongly indicating the universal behavior of vortex shedding dynamics in finite-temperature superfluid system.

Although there exists a notable resemblance in vortex shedding dynamics between classical and quantum systems, it is important to recognize a key distinction in the quantum context: the presence of multiple characteristic length scales. In quantum systems, the healing length, which characterizes the quantum vortex core, and the obstacle length scale both play essential roles. The interplay between these two scales raises a crucial question: does the observed dynamical similarity persist when the obstacle length scale approaches the same order of magnitude as the healing length? Furthermore, the extent to which a well-defined Reynolds number can be established in such situation remains uncertain. Current formulations of the quantum Reynolds number primarily incorporate the obstacle length as the characteristic scale, neglecting the potentially critical role of the healing length. This omission not only complicates the characterization of quantum fluid dynamics but also casts doubt on the applicability of established dynamical similarity principles in these regimes. Hence, a deeper investigation into the interplay of these length scales and their implications for dynamical similarity in quantum systems is needed and will be explored in our future work.

A notable distinction between our simulation and zero-temperature systems is the emergence of a static vortex dipole train, where complex interactions between the supercurrents of the vortices and the background occur, resulting in the compensation of the bulk supercurrent within the dipole train. This effect imbues the vortex dipole train with rigidity. However, in this work, we are constrained to analyzing the configuration of this vortex shedding pattern. Whether this configuration remains stable in the presence of larger background supercurrents remains an open question.

\section*{Acknowledgement}
S.L. is partially supported by the funding of Guangdong Basic and Applied Basic Research Foundation of China (Grant Nos. 2024A1515012552, 2022A1515011938). Y.T. is partially supported by the National Natural Science Foundation of China (under Grants No. 12035016, 12375058 and 12361141825). P.Y. acknowledge the support by the China Postdoctoral Science Foundation under Grant Number 2024T170545.
H.Z. is partly supported by the National Key Research and Development Program of China with Grant No. 2021YFC2203001 as well as the National Natural Science Foundation of China with Grant Nos. 12075026 and 12361141825.
\bibliography{biblio}

\appendix 

\newpage
\clearpage

\onecolumngrid %%%%%%%%%%%%%%%%%%%%%%%%%%%%%
\newpage           %%%%%%%%%%%%%%%%%%%%%%%%%%%%%

\section{Dynamical evolution in co-moving reference system}\label{comove}%%%%%%%%%
In the co-moving reference system, the equations of motion take the following form
\begin{eqnarray}
\partial_t \partial_z \phi&=&\partial_z(\frac{f(z)}{2}\partial_z\phi)+\frac{1}{2}\partial^2\phi-i\textbf{A}
\cdot\partial\phi +i A_t\partial_z\phi 
-\frac{i}{2}(\partial\cdot\textbf{A}-\partial_zA_t+v_x\partial_z A_x)\phi\\
&-&\frac{1}{2}(z+\textbf{A}^2)\phi+v_x \partial_x\partial_z\phi+i v_x A_x\partial_z\phi, \label{dphi_co}\nonumber\\
\partial_t \partial_z A_x&=&\partial_z(\frac{f(z)%-v_x^2
}{2}\partial_z A_x)-|\phi|^2 A_x+\textrm{Im}(\phi^*\partial_x\phi+v_x\phi^*\partial_z\phi)
+\frac{1}{2}(v_x\partial_z^2A_t\\
&+&\partial_x\partial_z A_t+\partial_y^2 A_x -\partial_x\partial_y  A_y-v_x\partial_y\partial_zA_y), \label{eqeom_x_co}\nonumber\\
\partial_t \partial_z A_y&=&\partial_z(\frac{f(z)}{2}\partial_zA_y)-|\phi|^2A_y+\textrm{Im}(\phi^*\partial_y\phi)
+\frac{1}{2}\left[\partial_y\partial_zA_t+\partial_x^2\textbf{A} -\partial_y\partial_x A_x-v_x\partial_y\partial_zA_x\right]\\
&+&v_x\partial_x\partial_zA_y, \label{eqeom_y_co}\nonumber \\
\partial_t\partial_z A_t&=&\partial^2A_t-\partial_t\partial\cdot\textbf{A}+f(z)\partial_z\partial\cdot\textbf{A} -2A_t|\phi|^2
+2\textrm{Im}(\phi^*\partial_t\phi)-2f(z)\textrm{Im}(\phi^*\partial_z\phi)\\
&-&v_x(A_x |\phi|^2+\textrm{Im}(\phi^*\partial_x\phi)+\frac{1}{2}(\partial_y^2A_x-\partial_x\partial_zA_t-\partial_z\partial_tA_x+\partial_x\partial_zA_x-\partial_x\partial_yA_y)).
\label{eqAt}\nonumber \\
\partial_z^2A_t&=&\partial_z\partial\cdot\textbf{A}+v_x\partial_z^2 A_x-2\textrm{Im}(\phi^*\partial_z\phi)
\end{eqnarray}

During the evolution, the presence of relative velocity between obstacle and normal fluid as well as emission of vortexs leading to the energy of bulk system is not conserved. Just like \cite{doi:10.1126/science.1233529}, one can derive the stress tensor 
\begin{equation}
    T_\nu^\mu=\frac{1}{2}{F_{\nu\alpha}F^{\mu\alpha}-\frac{1}{2}\delta^\mu_\nu(\frac{1}{2}F_{\alpha\beta}F^{\alpha\beta}+D_\alpha\Psi^*D^\alpha\Psi+m^2|\Psi|^2)+2D_\nu\Psi^*D^\mu\Psi},
\end{equation}
which satisfied the conservation equation as
\begin{equation}
    \partial_\mu\sqrt{-g}T_\nu^\mu=0.
\end{equation}
Therefore, the total energy $F$ that dissipated into black hole is related with $T^z_t$ and can be definied as
\begin{equation}
    F=-\int dx \sqrt{-g}T^z_t(z_H).
\end{equation}

\begin{figure}[htb]
\centering
\includegraphics[width=0.6\linewidth]{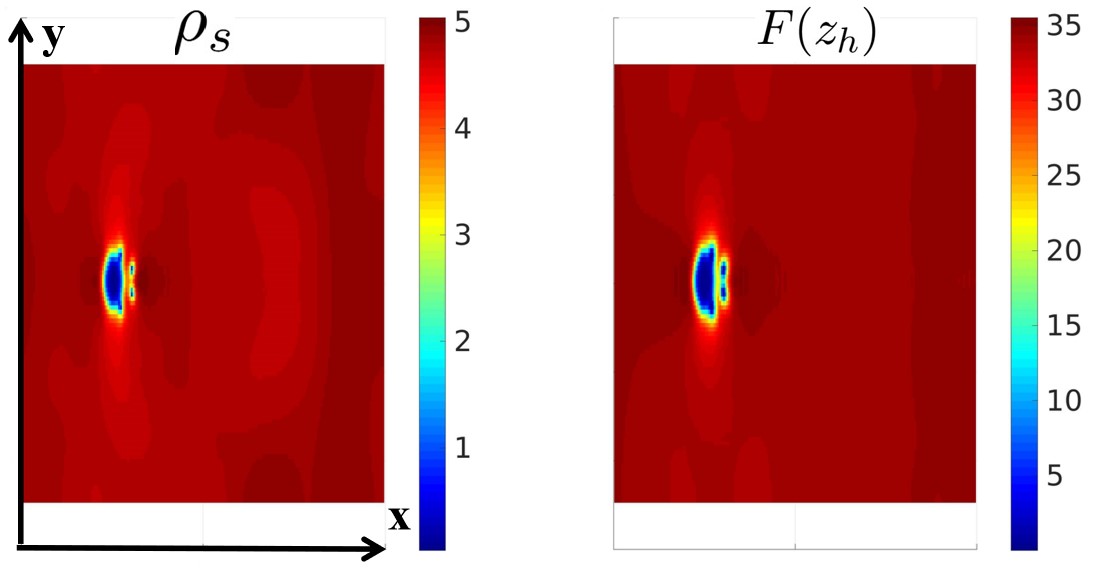}
\caption{Left: Superfluid condensate density $\rho_s$. Right: Energy dissipation density at horizon $F(z_h)$.}
\label{dissipation}
\end{figure}

\end{document}